# Improved Current Densities in MgB$_2$ By Liquid-Assisted Sintering

S. K. Chen, Z. Lockman, M. Wei, B. A. Glowacki

and J. L. MacManus-Driscoll

Department of Materials Science and Metallurgy, University of Cambridge,

Pembroke Street, Cambridge CB2 3QZ, UK

Polycrystalline MgB$_2$ samples with GaN additions were prepared by reaction of Mg, B, and GaN powders. The presence of Ga leads to a low melting eutectic phase which allowed liquid phase sintering and produces plate-like grains. For low-level GaN additions (≤ 5% at. %), the critical transition temperature, $T_c$, remained unchanged and in 1T magnetic field, the critical current density, $J_c$ was enhanced by a factor of 2 and 10, for temperatures of ~5K and 20K, respectively. The values obtained are approaching those of hot isostatically pressed samples.

MgB$_2$ is an intermediate temperature, type II superconductor with $T_c$ of ~ 40K, and with promising applications potential in high field magnets, motors and magnetic resonance imaging. It is now well known that high grain boundary angles do not produce weakly linked grain boundaries. Hence, instead of focusing on texture improvement (as for high temperature superconductors), materials engineering approaches have been directed towards either improving $J_c$ by improving grain connectivity, or improving in-field performance by alloying or addition of flux pinning centres.

Several hot pressing studies have been conducted on MgB$_2$, using either hot uniaxial or hot isostatic pressing (HIP) [1-6] and have shown improvements in $J_c$ by a factor of 6 at 5K, 4T [3]. For possible future MRI applications of MgB$_2$ which require fields near 2T to be generated and operation temperatures around 20-26K to be used, the improvements in $J_c$ through alloying are small, and it is the microstructure and sample density which dominate the current carrying capacity.

The aim of this work is to study the effect of the addition of Ga to Mg and B precursor powders on the microstructure and hence current carrying capacity of MgB$_2$ in low fields, of interest for MRI. Of course any enhancements in current carrying capacity at low fields will also be useful in alloyed samples of interest for higher field applications. Hence, the idea of this work is to replicate the promising results of hot pressing studies by using a simple ambient pressure reaction process. Ga forms a low melting (422°C) eutectic liquid with Mg and hence there is the possibility of achieving liquid phase sintering. We used GaN as reactant in the precursor powder as, unlike pure Ga or other salts of Ga, it is easy to handle and is not hygroscopic. In addition, it is known to decompose below 800 °C in the presence of H$_2$ gas [7], namely below the standard reaction temperature used for phase formation of MgB$_2$.

Doped MgB$_2$ pellets were prepared by in-situ reaction of Mg (Alfa Aesar, 99.6%), GaN (Aldrich, 99.99+%) and B (Alfa Aesar, 96-98%) powders. GaN of 0, 1, 3, and 5 mol % were added to the Mg and B powders. After manual grinding, the samples were uniaxially pressed at 10 ton/cm$^2$. Pellets (5 mm diameter x 2 mm thickness) were encapsulated in Ta foil (Advent, 99.9%) packed with powder of the same composition as the sample, and magnesium metal turnings. The envelope was then placed on a niobium-lined alumina boat. Reaction sintering was undertaken in a tubular furnace under a flowing 2% H$_2$-N$_2$ gas mixture. Samples were reacted at 950°C for 15 minutes using heating and cooling rates of 15°C/min.

X-ray diffractometry was conducted using Cu-K$\alpha$ radiation in the Bragg Bretano configuration to determine phase composition of the samples. Field emission gun scanning electron microscopy (FEG-SEM) was undertaken on fracture cross sections of samples to determine samples microstructures. Electron dispersive x-ray analysis (EDX) was also undertaken in the SEM. High resolution transmission electron microscopy (TEM) was undertaken on thinned pellet samples.

Superconducting transition temperatures were recorded using an AC Susceptometer. $J_c$ of the samples was determined using a D.C. SQUID Magnetometer. The sample (around 1 mm$^3$ in volume) was placed in a small capsule, and inserted into the magnetometer. The magnetometer was cooled to around 5 K using liquid helium. Measurements of magnetisation (*M*) versus magnetic field (*H*) were recorded on both whole pelletised, fully reacted samples. $J_c$ was determined using the Bean model and the whole sample dimension for the current length scale. Room temperature resistivity measurements were conducted in the Van der Pauw geometry. Density measurements were carried out by weighing the samples, and normalising by the sample geometric volume.

Figure 1 shows the Mg-Ga phase diagram in the composition and temperature region of interest to this study. A eutectic liquid is formed at 422°C, for 19.1 at.% Ga in Mg. Since we have GaN in our precursor mix, it is necessary first to decompose the GaN to Ga+$N_2$ before the liquid phase can form. According to phase stability studies of GaN in the presence of $H_2$, any significant decomposition will not occur below 700°C. Hence, the liquid phase will form in the same reaction temperature regime as the $MgB_2$ forms from the Mg and B precursors. If we assume that $MgB_2$ begins to form at as low as 550°C [9], then at just below this temperature, for a 3 at. % Ga sample Fig. 1 indicates that there is 10 mol. % liquid, 30 mol. % Mg, and 60 mol % B. As the temperature increases, the amount of liquid phase will increase but its percentage will depend on the rate of $MgB_2$ formation compared to heating rate.

Figure 2 compares x-ray diffraction patterns for 0 %, 1 at. % and 3 at. % GaN added $MgB_2$ samples. The pure and 1 at. % samples show clean x-ray patterns with peaks ascribed to $MgB_2$ only. The 3 at. % sample shows peaks from $Mg_5Ga_2$ (and possibly also $Mg_2Ga_5$ and $MgGa_2$) and Mg-B-N phases (e.g. $Mg_3BN_3$ and $Mg_3N_2$). Hence, the $N_2$ evolved from the GaN decomposition is not all transported away by the flowing gas, i.e. at least some of it reacts with Mg and B. From the x-ray patterns, there is no measurable preferential alignment of the $MgB_2$ grains in any of the samples. The 5 at. % sample (not shown) displays the same peaks as the 3 at. % sample, but there are larger quantities of second phase present.

Figure 3 compares FEGSEM fracture micrographs of a pure $MgB_2$ sample and of two different regions of a 3 at. % GaN sample (b and c). The grain morphologies of the pure (a) and doped (b) samples are strikingly different, with larger, more platey grains observed for the GaN doped samples. The appearance of a cracked surface-covering phase in c) with grains observed below the phase suggests the presence of a quenched, non-metallic liquid phase with partial surface oxidation. The regions covered ~ 20% of the fracture surface. Indeed, EDX indicated the composition of the surface layer to be $MgGa_{2-2.5}$. Fig. 3 (d) shows a high resolution image of a grain boundary region in a 3 at. % GaN sample. An amorphous region is apparent at the grain boundary. Again, these regions are common throughout the sample, and are much more numerous than in the pure sample. The amorphous phase is likely the same covering phase as seen in (c).

Table 1 shows $T_c$, room temperature resistivity, density, and grain size values for 0–5 at. % GaN doped samples. While $T_c$ does not change with GaN addition (indicative of no subsitution of Ga in the lattice), the resistivity shows a systematic increase from 28 µΩ.cm to 102 µΩ.cm. The resistivity of the pure sample is typical for the purities of the starting powders employed [10]. The sample density also increases systematically, although the density increase is small compared to what one typically achieves by hot pressing (~ 2.1 g/cm$^3$, which is ~ 80% of the theoretical value) [2]. The grain size increases from the 0% to 1% doped sample but changes little with further doping.

Figure 4 compares $J_c$ versus applied magnetic field up to 3T (// to the radial direction of the sample) for the whole pellets of the pure sample, the 3 at. % doped sample (the highest $J_c$ sample of all the samples prepared), a typical, clean uniaxially pressed $MgB_2$ sample from the literature [11] and a HIP'ed sample from the literature [4]. Fig. 4 (a) shows our data at 6K compared to others at 4.2K, and 5K. The two uniaxially pressed, pure $MgB_2$ samples show near identical behaviour. The HIP'ed sample has the highest $J_c$ over the whole field regime. $J_c$ (~5K, 1T) is 1x10$^5$ A.cm$^{-2}$, 2x10$^5$ A.cm$^{-2}$, and 4x10$^5$ A.cm$^{-2}$ for the uniaxial pure sample, 3 at. % GaN sample, and HIP'ed pure sample, respectively. Up to 2T, a field range of interest for MRI, the 3 at. % GaN sample shows approximately a factor of 2 improvement over the pure sample, and is approximately half the value of the HIP'ed sample. At 20K, the 3 at. % GaN sample shows a much stronger improvement in $J_c$ over the pure sample. $J_c$ (20K, 1T) is $3 \times 10^4$ A.cm$^{-2}$, $3 \times 10^5$ A.cm$^{-2}$, and $5 \times 10^5$ A.cm$^{-2}$ for the uniaxial pure sample, 3 at. % GaN sample, and HIP'ed pure sample, respectively. The field dependences of the samples are similar which is expected since none of the samples are alloyed.

For the HIP'ed sample, $J_c$ is improved through significant sample densification and grain connectivity. For the GaN added sample, the density is only marginally increased (by around 8% for the 3 at. % sample). The resistivity and second phase fraction both increase with GaN addition. In addition, TEM shows the presence of amorphous regions. These observations are consistent with obstruction of the current path by the second phases, and hence reduced connectivity. The improved $J_c$ must, therefore, originate from an increased intragrain $J_c$, i.e. improved carrying properties of the grains themselves. This is consistent with the observation of grains with a highly crystalline, faceted nature which are formed through liquid phase sintering. Hence, even though the current carrying cross section is reduced, the overall $J_c$ is still higher because of the enhanced intragranular properties. Atttempts to measure intragrain $J_c$ by forming finely crushed powders were not successful for these samples because the finest powder particles obtained (few micron) were more than an order of magnitude larger than the grain size.

In summary, liquid phase sintering has been used to enhance the current carrying properties of bulk MgB$_2$. The technological significance of improving current carrying capacity without using expensive and size- limiting hot pressing equipment is clear[11].

Table 1    $T_c$, room temperature resistivity, density and grain size of pure and doped samples.

| GaN doped sample | $T_c$ (K) | Resistivity (290K), $\rho$ (µΩ.cm) | Density (g.cm$^{-3}$) | Grain size (µm) |
|---|---|---|---|---|
| 0% | 38.9 | 27.8 | 1.16 | 0.13 – 0.44 |
| 1% | 38.9 | 56.2 | 1.22 | 0.53 – 0.96 |
| 3% | 38.8 | 61.1 | 1.25 | 0.45 – 0.87 |
| 5% | 38.8 | 102.1 | 1.31 | 0.35 – 0.65 |

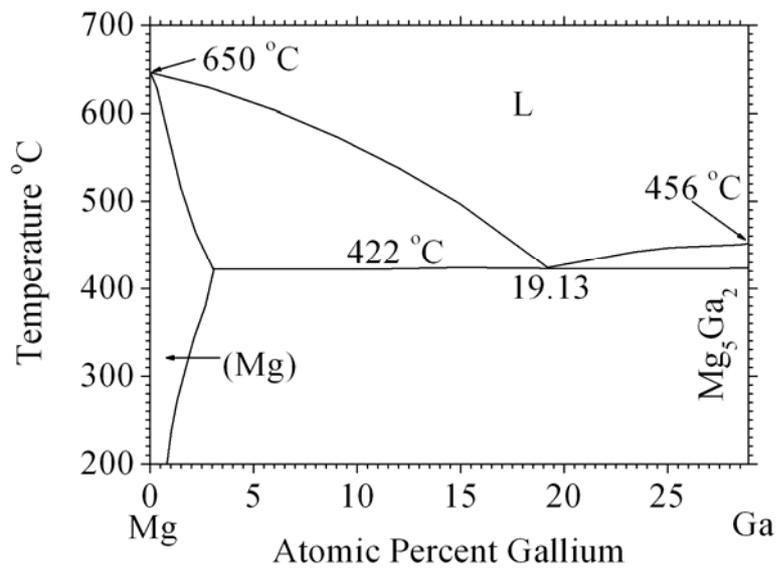

FIG. 1. Mg-Ga phase diagram (after T. Massalski, Ref. 8).

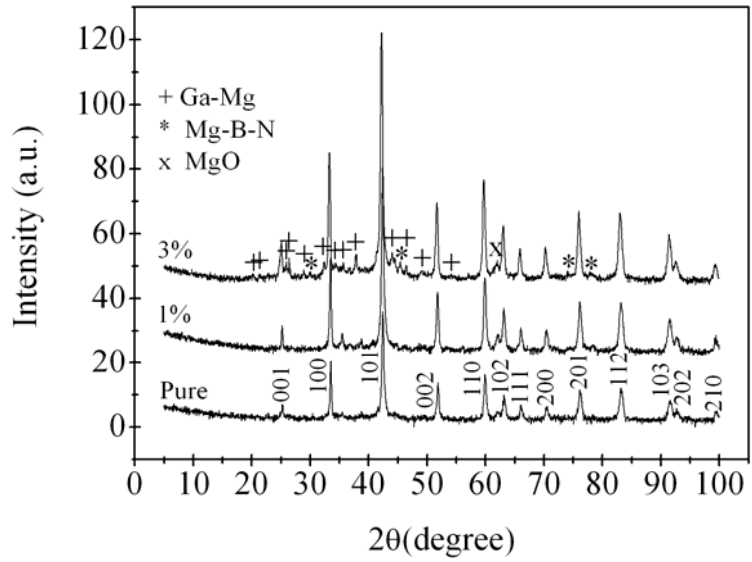

FIG. 2. X-ray diffraction patterns of the pure and GaN doped samples.

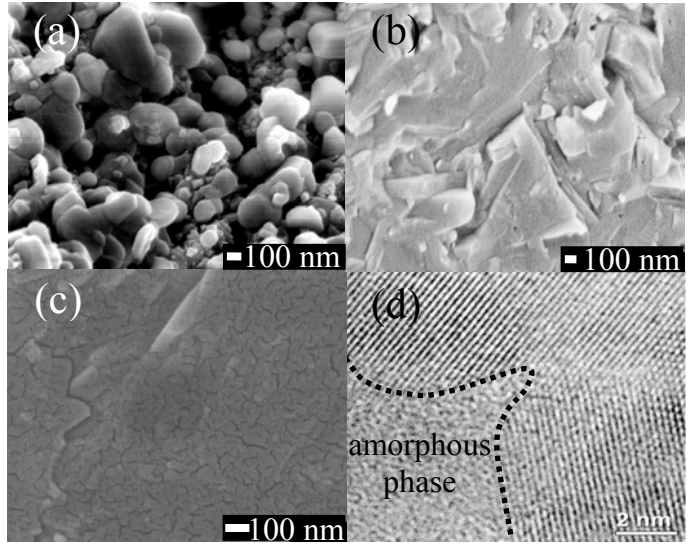

FIG. 3. SEM images of (a) pure $MgB_2$, b & c) 3 at.% GaN doped samples at different magnification and (d) TEM image of a grain boundary region in the 3 at. % GaN doped sample.

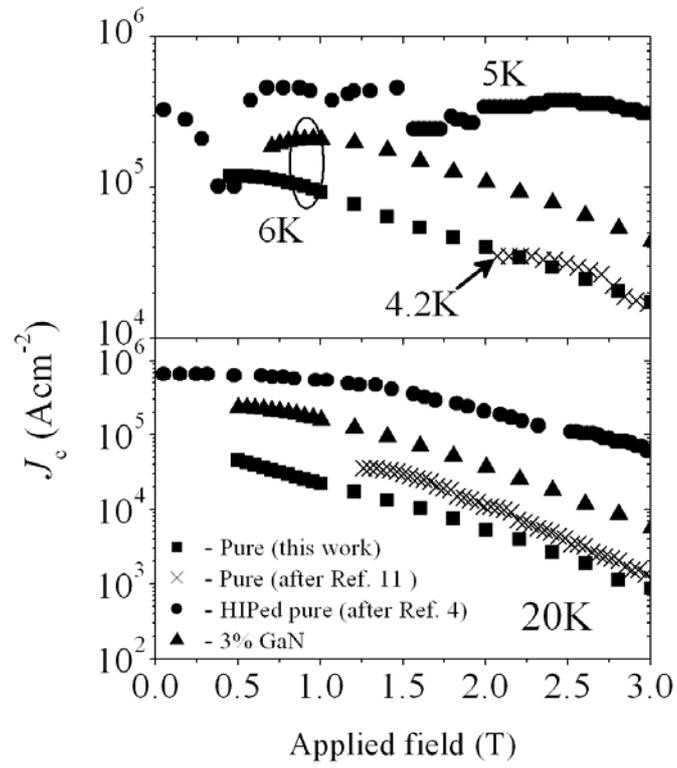

FIG. 4. Comparison of $J_c$ versus applied magnetic field for samples in this work and from literature (Ref. 4); (Ref. 11).